\input{aipcheck}

\documentclass[
    ,final            
  ]
  {aipproc}

\layoutstyle{6x9}

\usepackage{epsfig}

\begin{document}

\title{The $\Theta^+$ (1540)
as a heptaquark with the overlap of a pion, a kaon and a nucleon}
\author{P. Bicudo}{address={
Dep. F\'{\i}sica and CFIF, Instituto Superior T\'ecnico, Av.
Rovisco Pais
1049-001 Lisboa, Portugal}}
\author{G. M. Marques}{address={
Dep. F\'{\i}sica and CFIF, Instituto Superior T\'ecnico, Av.
Rovisco Pais
1049-001 Lisboa, Portugal}}

\begin{abstract}
We study the very recently discovered $\Theta^+$ (1540) at
SPring-8, at ITEP and at CLAS-Thomas Jefferson Lab. We apply the
same RGM techniques that already explained with success the
repulsive hard core of nucleon-nucleon, kaon-nucleon exotic
scattering, and the attractive hard core present in pion-nucleon
and pion-pion non-exotic scattering. We find that the $K-N$ repulsion
excludes the $\Theta^+$ as a $K-N$ s-wave pentaquark. We explore the
$\Theta^+$ as heptaquark, equivalent to a $N+\pi+K$ borromean boundstate,
with positive parity and total isospin $I=0$. We find that the kaon-nucleon
repulsion is cancelled by the attraction existing both in the pion-nucleon
and pion-kaon channels. Although we are not yet able to bind the total
three body system, we find that the $\Theta^+$ may still be a heptaquark state.
\end{abstract}
\maketitle


\par
In this talk we study the exotic hadron $\Theta^+$ 
\cite{update}
(narrow hadron resonance of 1540 MeV decaying into a $n K^+$) 
very recently discovered at Spring-8 
\cite{Spring-8}, 
and confirmed by ITEP 
\cite{ITEP} 
and by CLAS at the TJNL
\cite{CEBAF}. 
This is an extremely exciting state because it may
be the first exotic hadron to be discovered, with quantum numbers
that cannot be interpreted as a quark and an anti-quark meson or
as a three quark baryon. 
Very recent studies propose that the $\Theta^+$ is a pentaquark state
\cite{recent}. 
Exotic multiquarks are expected since the early works of Jaffe 
\cite{Jaffe,Strottman},
and some years ago Diakonov, Petrov and Polyakov
\cite{Diakonov}
applied skyrmions to a precise prediction of the $\Theta^+$.
The nature of this particle, and its isospin, 
parity 
\cite{Hyodo} 
and angular momentum are yet to be determined.

\par
\begin{figure}[t]
\begin{picture}(100,170)(0,0)
\multiput(133,0)(0,2){76}{$\cdot$}
\put(-69.75,1.75){\epsfig{file=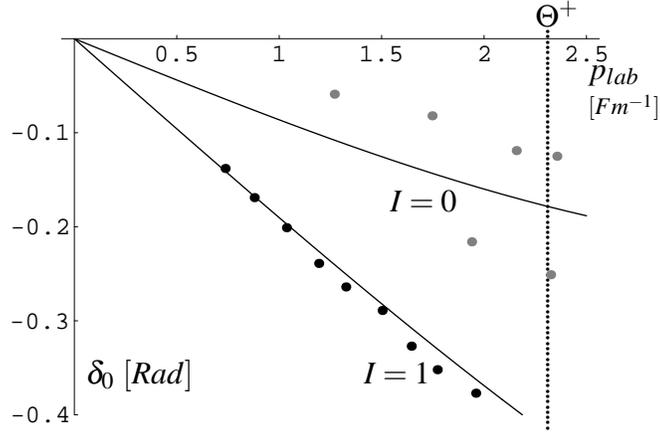,width=8cm}}
\put(150,135){$ p_{lab} $} \put(150,125){$ _{[Fm^{-1}]} $}
\put(-40,20){$\delta_0  \ [Rad] $} \put(75,85){$ I=0$}
\put(65,20){$ I=1$} \put(130,155){$ \Theta^+$}
\end{picture}
\caption{The $I=0$ and $I=1$ experimental \cite{Barnes} and
theoretical (this talk and ref. \cite{Bicudo}) s-wave phase
shifts as a function of the kaon momentum in the laboratory
frame.} \label{knexp}
\end{figure}

We start in this talk by reviewing the Quark Model (QM) and the
Resonating Group Method (RGM) \cite{Wheeler}, which are adequate to 
study states where several quarks overlap. We show that the QM, together
with chiral symmetry, produces hard core hadron-hadron potentials,
which can be either repulsive or attractive. First we apply the
RGM to show \cite{Bicudo,Bender} that the exotic $N-K$ hard core
s-wave interaction is repulsive. This is consistent with the
experimental data \cite{Barnes}, see Fig. \ref{knexp}. We think
that this excludes the $\Theta^+$ as a bare s-wave pentaquark
$uddu\bar s$ state or a tightly bound s-wave $N - K$ narrow
resonance. The observed mass of the $\Theta^+$ is larger than 
the sum of the $K$ and $N$ masses by $1540-939-493=118 \ MeV$,
and this does not suggest a simple $K-N$ binding. However a $\pi$ 
could also be present in this system, in which case the binding 
energy would be of the order of $20 \ MeV$.
Moreover this state of seven quarks would have a positive parity,
and would have to decay to a p-wave $N - K$ system, which is
suppressed by angular momentum, thus explaining the narrow width
of the $\Theta^+$. With this natural description in mind we then
apply the RGM to show that the $\pi - N$ and $\pi - K$ hard core
interactions are attractive. Finally we put together the $\pi -
N$, $\pi - K$ and $N-K$ interactions to show that the $\Theta^+$
is possibly a borromean \cite{borromean} three body s-wave
bound state of a $\pi$, a $N$ and a $K$, with positive parity and
total isospin $I=0$.

%
%

\par
The RGM was used by Ribeiro \cite{Ribeiro} to show that in exotic
hadron-hadron scattering, the quark-quark potential together with
Pauli repulsion of quarks produces a repulsive short range
interaction. For instance this explains the $N - N$ hard core
repulsion, preventing nuclear collapse. Deus and Ribeiro
\cite{Deus} used the same RGM to show that, in non-exotic
channels, the quark-antiquark annihilation could produce a short
core attraction. The RGM computes the effective hadron-hadron
interaction using the matrix elements of the microscopic
quark-quark interactions. The wave functions of quarks are
arranged in anti-symmetrized overlaps of simple colour singlet
hadrons. 

When spontaneous chiral symmetry breaking is
included in the quark model \cite{Bicudo3,Bicudo4} the
annihilation potential becomes crucial \cite{Bicudo1,Bicudo2} to
understand the low mass of the $\pi$. The annihilation potential
$A$ is also present in the $\pi$ Salpeter equation where it
cancels most of the kinetic energy and confining potential $2T+V$.
From the quark model with chiral symmetry breaking we get the matrix
element \cite{Bicudo3},
\begin{eqnarray}
\langle A \rangle_{S=0} &=& - {2 \over 3} (2M_N-M_\Delta) \ .
\label{sum rules}
\end{eqnarray}
While the standard quark-quark potential provides a repulsive (positive) 
overlap, it is clear in eq. (\ref{sum rules}) that the annihilation
potential provides an attractive (negative) overlap. This
confirms that the hard core can be attractive for non-exotic
channels where annihilation occurs, while it is repulsive for 
exotic channels.

We summarize
\cite{Bicudo,Bicudo1,Bicudo2} the effective potentials 
for the different channels,
\begin{eqnarray}
V_{K-N}&=& \frac{2-\frac{4}{3}  \vec \tau_A \cdot \vec
\tau_B}{\frac{5}{4}+\frac{1}{3}  \vec \tau_A \cdot \vec \tau_B}
\frac{(M_\Delta-M_N )}{3} \left( 2 \sqrt{\pi} \over \alpha \right)^3 
e^{- \frac{{p_\lambda}^2}{2\beta^2
}} \int \frac{d^3 p'_\lambda}{(2 \pi)^3} \, e^{- \frac{
{p'_\lambda}^2}{2 \beta^2}}
\nonumber \\
V_{\pi-N}&=& \frac{2}{9}( 2M_N - M_\Delta ) \,  \vec \tau_A \cdot
\vec \tau_B \, {\cal N_\alpha}^{-2} \ ,
\nonumber \\
V_{\pi-K}&=& \frac{8}{27}(2 M_N - M_\Delta) \, \vec \tau_A \cdot
\vec \tau_B \, {\cal N_\alpha}^{-2} \ , \label{zero p}
\end{eqnarray}
where $\vec \tau$ are the isospin matrices. This parametrization
in a separable potential enables us to use standard techniques
\cite{Bicudo} to exactly compute the scattering $T$ matrix. The
scattering length $d \delta_0 / dk_{c.m.}$ is,
\begin{equation}
a=-{ \sqrt{\pi} \over \alpha} { 4 \mu \, v \over \alpha ^2 +
{\beta \over \alpha} 4 \,  \mu \, v } \ .
\label{teor scatt}
\end{equation}
The parameters and results for the relevant channels are
summarized in Table \ref{scattering lengths}. We have fitted
$\alpha$ with the $I=1$ $K-N$ scattering. We use $\beta=\alpha$ in
the repulsive channels and fit it with the appropriate scattering
lengths in the attractive $I=1/2$ pionic channels.
%
%
\begin{table}[t]
\begin{tabular}{c|cccccc}
channel                   & $\mu$  & $v_{th}$&$\alpha$ &$\beta$& $a_{th}$& $a _{exp}$ \\
\hline
$  K-N_{ I=0 }         $ & $ 1.65$ & $ 0.50$ & $ 3.2$ & $ 3.2$ & $-0.14$ & $ -0.13\pm 0.04 $
\cite{Barnes} \\
$  K-N_{ I=1 }         $ & $ 1.65$ & $ 1.75$ & $ 3.2$ & $ 3.2$ & $-0.30$ & $ -0.31\pm 0.01 $
\cite{Barnes} \\
$ \pi-N_{I={1\over 2}} $ & $ 0.61$ & $-0.73$ & $ 3.2$ & $ 11.4$ & $ 0.25$ & $  0.246\pm 0.007$
\cite{Itzykson} \\
$ \pi-N_{I={3\over 2}} $ & $ 0.61$ & $ 0.36$ & $ 3.2$ & $ 3.2$ & $-0.05$ & $ -0.127\pm 0.006$
\cite{Itzykson} \\
$ \pi-K_{I={1\over 2}} $ & $ 0.55$ & $-0.97$ & $ 3.2$ & $ 10.3$ & $ 0.35$ & $  0.27\pm 0.08 $
\cite{Nemenov} \\
$ \pi-K_{I={3\over 2}} $ & $ 0.55$ & $ 0.49$ & $ 3.2$ & $ 3.2$ & $-0.06$ & $ -0.13\pm 0.06 $
\cite{Nemenov} \\
\hline
\end{tabular}
\caption{ This table summarizes the parameters $\mu , \, v \,
,\alpha \, , \beta$ (in Fm$^{-1}$)
 and scattering lengths $a$ (in Fm) .}
\label{scattering lengths}
\end{table}

\par
Let us first apply our method to the $K-N$ exotic system, where
the anti-quark $\bar s$ is present. In this case the $I=0$ channel
is less repulsive than the $I=1$ channel. With our method we
reproduce the $K-N$ exotic s-wave phase shifts -- see Fig.
\ref{knexp} -- where indeed there is no evidence for the
$\Theta^+$ state. In what concerns the $\pi - N$ system and the
$\pi - K$ systems, the corresponding parameters in Table
\ref{scattering lengths} are almost identical. There we find
repulsion for $I=3/2$ and attraction for $I=1/2$. The repulsion in
the $I=3/2$ channel prevents a bound state in this channel. In
what concerns the $I=1/2$ channel, the attraction is not
sufficient to provide for a bound state, because the $\pi$ is
quite light and the attractive potential is narrow. From eq.
(\ref{teor scatt}) we conclude that we have binding if the reduced
mass is $\mu \geq -{\alpha^3 \over 4 \beta v }$. With the present
parameters $\mu$ should be of the order of 1 Fm$^{-1}$. This
is larger than the $\pi$ mass, therefore it is not possible to
bind the $\pi$ to the $K$ or to the $N$. All that we can get is a
very broad resonance. For instance in the $\pi - K$ channel this
is the kappa resonance \cite{Rupp1}, which has been recently
confirmed by the scientific community. However, with a doubling of
the interaction, produced by a $K$ and a $N$, we expect the $\pi$
to bind.

\par
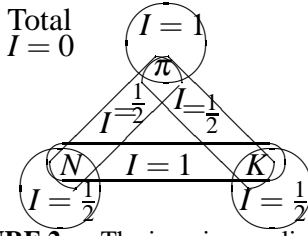
\begin{figure}[t]
%
\begin{picture}(200,70)(0,0)
\put(20,5){
\begin{picture}(120,70)(0,0)
\put(20,0){
\begin{picture}(100,100)(0,0)
\put(0,0){\oval(30,30)}
\put(0,5){$N$}
\put(-12,-8){$I={1\over 2}$}
\put(40,55){\oval(30,30)}
\put(35,43){$\pi$}
\put(30,58){$I=1$}
\put(80,0){\oval(30,30)}
\put(70,5){$K$}
\put(68,-8){$I={1\over 2}$}
\end{picture}}
\put(20,0){
\begin{picture}(20,100)(0,0)
\put(40,10){\oval(90,14)}
\put(25,5){$I=1$}
\put(41,40){\oval(20,20)[tl]}
\put(31,40){\line(1,-1){40}}
\put(41,50){\line(1,-1){40}}
\put(71,10){\oval(20,20)[br]}
\put(43,30){$I$}
\put(46,27){$=$}
\put(54,25){$1 \over 2$}
\put(6,10){\oval(20,20)[bl]}
\put(-4,10){\line(1,1){40}}
\put(5,-1){\line(1,1){40}}
\put(36,40){\oval(20,20)[tr]}
\put(15,21){$I$}
\put(20,26){$=$}
\put(26,29){$1 \over 2$}
\end{picture}}
\put(20,0){
\begin{picture}(20,100)(0,0)
\put(-20,60){Total}
\put(-20,50){$I=0$}
\end{picture}}
\end{picture}}
\end{picture}
%
\caption{ The isospin couplings in the $Z/\Theta$. } \label{ISO 1540}
\label{isospin}
\end{figure}
\begin{figure}[b]
\begin{picture}(100,120)(0,0)
\put(-70,0){\epsfig{file=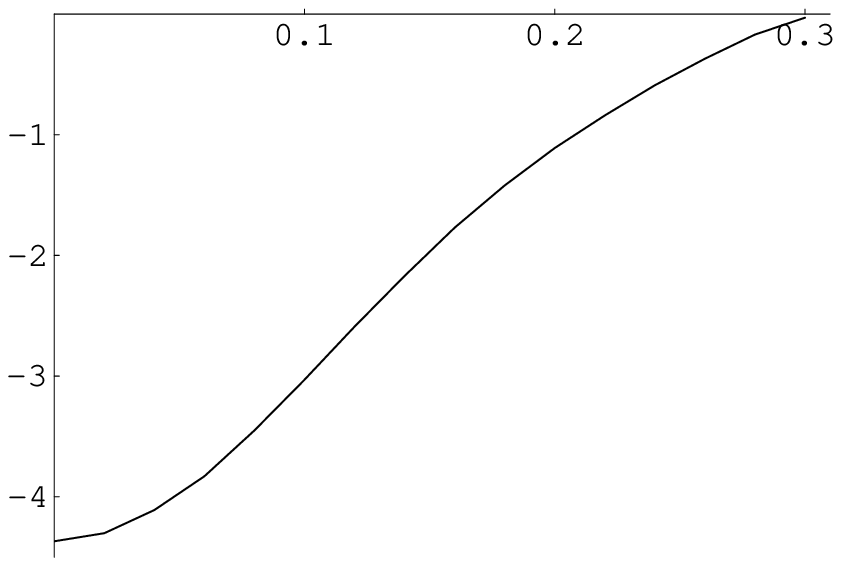,width=6.5cm}}
\put(110,100){$ 2a \ [Fm] $}
\put(0,20){$E_\pi (2a) \ [Fm^{-1}] $}
\end{picture}
\caption{ $\pi$ energy as a function of the
coordinate $ |\vec r_N-\vec r_K|=2a$.}
\label{pion energy}
\end{figure}

We now investigate the borromean \cite{borromean} binding of the
exotic $\Theta^+$ constituted by a $N$, $K$ and $\pi$ triplet. In
what concerns isospin, we need the $\pi$ to couple to both $N$ and
$K$ in $I=1/2$ states for attraction, and the only candidate for 
binding is the total $I=0$ state, see Fig. \ref{isospin}.
%
%
%
Since the $\pi$ is much lighter that the $N - K$ system, we study the
borromean binding adiabatically. As a first step, we start by
assuming that the $K$ and $N$ are essentially stopped and
separated by $\vec r_N-\vec r_K= 2a \hat e _z$. This will be improved
later. We also take advantage of the
similarities in Table \ref{scattering lengths} to assume that the
two heavier partners of the $\pi$ have a similar mass of 3.64
Fm$^{-1}$ and interact with the $\pi$ with the same separable
potential. Then we solve the bound state equation for a $\pi$ in
the potential $V_{\pi-N}+ V_{\pi-K}$,
where this potential is wider the direction of the $K - N$ $z$
axis. The $\pi$ energy is
depicted if Fig. \ref{pion energy}, and it is negative as
expected.
Once the $\pi$ binding energy is determined, we include it in the
potential energy of the $K - N$ system, which becomes the sum of
the repulsive $K - N$ potential and the $\pi$ energy. We find that
for short distances the total potential is indeed attractive.
Finally, using this $K - N$ potential energy we solve the
schr\"odinger equation for the system, thus including the
previously neglected $K$ and $N$ kinetic energies. 
However here we are not able to bind the $K-N$
system, because the total effective $K-N$ potential is not
sufficiently attractive to cancel the positive $K-N$ kinetic
energy. Nevertheless the $K$ is heavier than the $\pi$, thus
a small enhancement of the attraction would suffice to bind the
heptaquark. We remark that existing examples of narrow resonances 
with a trapped $K$ are the $f_0(980)$ 
\cite{Rupp1}, 
the $D_s(2320)$ 
\cite{Rupp2} 
very
recently discovered at BABAR 
\cite{Babar}, 
and possibly the $\Lambda(1405)$. 
Moreover we expect that meson exchange interactions
and the irreducible three-body overlap of the three hadrons, that
we did not include here, would further increase the attractive
potential. Therefore it is plausible that a complete computation
will eventually bind the $K-\pi-N$ system.
%

\par
We conclude that the $\Theta^+$ hadron very recently discovered
cannot be an s-wave pentaquark. We also find that it may be a
heptaquark state, composed by the overlap of a $\pi$, a $K$ and a
nucleon. This scenario has many interesting features. The $\Theta^+$ 
would be, so far, the only hadron with a trapped $\pi$. Moreover the 
$\pi$ would be trapped by a rare three body borromean effect. 
And the decay rate to a $K$ and a $N$ would be suppressed since 
the $\pi$ needs to be absorbed with a derivative coupling, while
the involved hadrons have a very low momentum in this state. 
Because the $\Theta^+$ would be composed by a $N$ and two 
pseudoscalar mesons, its parity would be positive, $J^P={1\over2}^+$, 
in agreement with the prediction of Diakonov, Petrov and Polyakov. 
The isospin of the $\Theta^+$ would be $I=0$ in order to ensure the 
attraction of the $\pi$ both by the $N$ and the $K$.

\begin{theacknowledgments}
We are very grateful to George Rupp for pointing our attention to the
pentaquark state.
The work of G. Marques is supported by Funda\c c\~ao para a
Ci\^encia e a Tecnologia under the grant SFRH/BD/984/2000.
\end{theacknowledgments}


\end{document}